\documentstyle[aps,prl,preprint]{revtex}
\def\la{\mathrel{\mathpalette\fun <}}

\def\fun#1#2{\lower3.6pt\vbox{\baselineskip0pt\lineskip.9pt
        \ialign{$\mathsurround=0pt#1\hfill##\hfil$\crcr#2\crcr\sim\crcr}}}
\def\l{\lambda}
\def\n{n(R,t)}
\def\f{\phi}
\begin{document}
\vspace*{-62pt}
\begin{flushright}
DART-HEP-94/02\\
May 1994
\end{flushright}

\vspace{0.75in}
\centerline{\bf ON THE STRENGTH OF FIRST ORDER PHASE TRANSITIONS}
\vskip 1.cm
\centerline{Marcelo Gleiser}

\vskip .5 cm
\centerline{Department of Physics and Astronomy}
\centerline{Dartmouth College}
\centerline{Hanover, NH 03755, USA}

\def\mpl{{m_{Pl}}}
\def\x{{\bf x}}
\def\p{\phi}
\def\F{\Phi}
\def\s{\sigma}
\def\a{\alpha}
\def\d{\delta}
\def\t{\tau}
\def\r{\rho}
\def\beq{\begin{equation}}
\def\eeq{\end{equation}}
\def\ba{\begin{eqnarray}}
\def\ea{\end{eqnarray}}
\def\re#1{^{\ref{#1}}}

\baselineskip 16pt
\vskip .5 cm
\baselineskip 10pt
Electroweak baryogenesis may solve one of the most fundamental questions
we can ask about the universe, that of the origin of matter. It has become
clear in the past few years that  it also poses a multi-faceted challenge. In
order to compute the tiny primordial baryonic excess, we probably must invoke
physics beyond the standard model (an exciting prospect for most people),
we must push perturbation theory
to its ``limits'' (or beyond),
and we must deal with nonequilibrium aspects of the
phase transition. In this talk, I focus mainly on the latter issue,
that of nonequilibrium aspects of first order transitions. In particular,
I discuss the elusive question of ``weakness''. What does it mean to have
a weak first order transition, and how can we distinguish between weak and
strong? I argue that weak and strong transitions have very different dynamics;
while strong transitions proceed by the usual bubble nucleation mechanism, weak
transitions are characterized by a mixing of phases as the system reaches the
critical temperature from above. I show that it is possible to clearly
distinguish between the two, and discuss consequences for studies of first
order transitions in general. \\
(Invited talk given at the ``Electroweak Physics and the Early Universe''
workshop, Sintra, March 23-25, 1994.)


\baselineskip 16pt
\def\beq{\begin{equation}}
\def\eeq{\end{equation}}
\def\ba{\begin{eqnarray}}
\def\ea{\end{eqnarray}}
\def\re#1{[{\ref{#1}]}}

\def\mpl{{m_{Pl}}}
\def\x{{\bf x}}
\def\p{\phi}
\def\F{\Phi}
\def\s{\sigma}
\def\a{\alpha}
\def\d{\delta}
\def\t{\tau}
\def\r{\rho}

\vspace{16pt}
\noindent{\bf I. INTRODUCTION}\\

Since the pioneering work of Becker and D\"oring on the nucleation of droplets
in fluids \re{BD}, the theoretical
investigation of nonequilibrium aspects of first order phase transitions has
been of interest to researchers in subjects ranging from
metereology and materials science to quantum field theory and cosmology.
Phenomenological field theories were developed by Cahn and Hilliard
and by Langer \re{LANGER} in the context of  coarse-grained time-dependent
Ginsburg-Landau models, in which an expression for the decay rate of the
metastable stable state was obtained within a steady-state formulation \re{CH}.
The study of metastable decay in zero-temperature quantum field theory
was initiated by Voloshin, Kobzarev and Okun \re{VKO} and soon after
put onto firmer theoretical ground by Coleman and Callan \re{COLEMAN}. Finite
temperature vacuum decay was first analysed by Linde \re{LINDE} and has
received considerable attention in the recent literature \re{FINITETDECAY}.

The main motivation for studying metastable decay for quantum fields comes from
the possibility that the Universe underwent a series of phase transitions
as it expanded from its initial hot, and presumably highly symmetric,
state \re{CPT}. These cosmological phase
transitions have the potential of not only answering many questions left open
by the standard big-bang model, but also of serving as windows to high energy
physics unnaccessible to  current (and future) terrestrial accelerators.
However, as it has become increasingly clear during the past few years,
in order to reliably compute quantities of interest, a deeper understanding
of the nonequilibrium aspects of the transitions is needed. And we all
know that nonequilibrium statistical mechanics is a hard subject; most of the
nice universal properties that appear in equilibrium statistical mechanics
are lost, and we are forced to study how nonlinearities will influence
the approach to equilibrium of this or that particular system. But things are
not all that bad. The very fact that we cannot but embark into
the study of nonequilibrium dynamics in the context of finite temperature
relativistic quantum field theories is also a blessing; as decades of
research on the dynamics of phase transitions in condensed matter systems
have shown, the subject is extremely rich, offering a wealth of interesting
possibilities. It is tempting to speculate that this will also be the case
for cosmological phase transitions and that new, unsuspected phenomena are
lurking behind our present level of understanding.

This talk is organized as follows. In the next Section I briefly discuss
some of the problems related with the weakness of the electroweak potential,
such as its evaluation and possible scenarios for the dynamics of the
transition. In Section 3 I present the results of
a numerical experiment in which a
very clear criterion to distinguish between ``weak'' and strong first order
transitions emerges. In Section 4 I briefly review some of the work on
subcritical bubbles emphasizing the
qualitative agreement between the subcritical
bubbles picture and the results of the numerical experiment of Section 3. In
Section 5 I present some concluding remarks.

\vspace {0.5cm}

\noindent{\bf II. The Issue: The Weakness of the Electroweak Transition}\\

Of the many interesting possibilities raised by primordial phase transitions,
the generation of the baryon number of the Universe during the electroweak
phase transition has been explored extensively since the seminal work
of Kuzmin, Rubakov, and Shaposhnikov \re{KRS}. For the purpose of this talk,
the important aspect of the electroweak phase transition is that it is, in
most scenarios proposed so far (see, e.g. \re{BRANDEN} for an alternative
approach), a first order
phase transition. And, at least within the context of the standard model of
particle physics, the transition is very possibly a weak one; the standard
computation for nucleation of critical bubbles shows that the thin-wall
approximation fails and that the bubbles are rather thick \re{THICK}. In
fact, this seems to be the case even for extensions of the standard model,
which according to some authors offer better hope of producing the correct
baryon asymmetry. (Contrast, for example, the contributions of Shaposhnikov
and Turok in these Proceedings.)

The weakness of the transition poses a tremendous challenge even to the
study of equilibrium properties of the system; improving the
perturbative evaluation of the effective potential proves to be a very
demanding and ungrateful task, as technical difficulties are compounded by
the fact that nonperturbative effects must be called for to regulate
the perturbative expansion near the symmetric phase \re{EWPERTURB}. Using
alternative methods such as the $\epsilon$-expansion offer an interesting
posibility, which, nevertheless relies on the success that these methods have
on different systems \re{EPSILON}. Another alternative is to go
to the computer and study the equilibrium properties of the
standard model on the lattice \re{EWLATTICE}.
Recent results are encouraging inasmuch as they seem to be consistent with
perturbative results in the broken phase
for fairly small Higgs masses. The transition also seems to be stronger
than the perturbative estimates would predict.
For larger Higgs masses (for this author, smaller than the 80 GeV claimed in
Ref. \ref{EWLATTICE})
the interpretation of the results is not very straight forward, as finite-size
effects become more important, and distinguishing the two phases by the
double-peak structure of the distribution function becomes trickier.

It is clear from the above paragraph that much work must be done before
we can claim we understand the equilibrium properties of the electroweak phase
transition in the context of the standard model.
Given that we do not know the exact shape of the effective potential
for realistic
Higgs masses, quantities such as the curvature of the potential around the
symmetric minimum,
the critical temperature of the transition, and maybe even the order of the
transition are still unknown. Even if one goes beyond the standard model,
as most people
prefer, some of these problems persist.

However, one thing seems to be certain;
that the transition is weakly first order. What does this mean exactly?
``A weak first order phase transition...'' In the context of electroweak
baryogenesis, the usage of the term {\it weak} to characterize the transition
is usually identified with the wall thickness of the nucleating bubbles.
A weak transition
would have typically thick bubbles, in that their radius is not much larger
than the wall's thickness. (Of course, in this case the definition of radius
is somewhat blurred.) On the other hand, the transition cannot be too weak
or not enough baryon number is generated. This is equated with the
discontinuity in the magnitude of the Higgs field at the critical temperature,
$\langle \phi (T_c) \rangle/T \la 1.$ As I hope will be clear later on,
this definition of weak, or not too weak, is not enough. Looking at the
effective potential (assuming you know it), identifying a barrier between
the symmetric and broken-symmetric phase, and proceeding to use the nucleation
rate formula for critical bubbles is not necessarily the right thing to do.
The reason for
this is simple; the vacuum decay formula is obtained by assuming that there
is a nicely behaved, nearly-homogeneous background about which we expand
the partition function in order to obtain the semi-classical approximation.
The reader is referred to Ref. \ref{GR} for details. However,
for small enough barriers between the two phases, large amplitude
fluctuations will
be present, invalidating the assumption of a near-homogeneous background.
Homogeneous nucleation breaks down for these situations. Two questions
immediately come to mind. When does homogeneous nucleation break down,
and what should one
do when it does? (Getting drunk on Port is not a very good long term solution,
posing problems immediately after you wake up the next day, {\it if} you wake
up the next day.)

With these questions in mind, and motivated by the ``weakness'' of the
electroweak transition, a few years ago
Gleiser and Kolb (GK) proposed a novel mechanism by
which such transitions
evolve \re{GK}. Rather than having nucleation of thick bubbles below
the critical
temperature of the transition, the transition would be characterized by
a substantial phase mixing as the critical temperature is approached from
above.
GK modelled the dynamics of this phase mixing by estimating
the fraction of the volume occupied by subcritical (correlation volume)
thermal fluctuations of
each phase at the critical temperature. In their initial
approach, they neglected the fact that these subcritical fluctuations were
unstable and thus were criticized on the grounds that they overestimated the
equilibrium fraction of the volume in the broken-symmetric phase \re{CRITICS}.
In their simple analytical approach the equilibrium fraction of both phases
should be an identical 0.5 each when the two phases are degenerate (at $T_c$).
Their method was recently extended by the authors of
Ref. \re{GG} to
include the shrinking of the bubbles. The results of Ref. \re{GG} indicate that
GK are at least qualitatively correct; there {\bf will}
be a regime in which the
transition is
weak enough that considerable phase mixing occurs above $T_c$. (It is, of
course, possible that this interesting regime lies beyond the validity of the
perturbative evaluation of the electroweak effective potential.
Presently this does not appear to be so \re{GG}.) I will briefly review
the subcritical bubbles method later on. Now I would rather proceed with
the discussion of the strength of the transition.

\vspace{0.5cm}

\noindent{\bf III. Distinguishing the Weak from the Strong: A Simple Model}\\

In order to sharpen the distinction between weak and strong first order
transitions, I decided to investigate this question within the context of
a simple toy model in 2+1 dimensions
which could be studied numerically \re{WEAK}.
Due to the complex nonequilibrium nature of the system, any analytical approach
(at least  those proposed so far) is bound to be severely limited.
One is justified in regarding these simple
models with suspicion. The need for a numerical investigation of this question
is clear. This need is even more justified by noting that several of the
{\it gross} features
of the electroweak transition may appear in other unrelated physical systems,
such as nematic liquid crystals and certain magnetic materials below their
critical temperature. Moreover, numerical
simulations of first-order transitions in the context of field theories
(as opposed to discrete Ising models \re{NUMISING}) are scarse.
Recent work has shown that
the effective nucleation barrier is accurately predicted by homogeneous
nucleation
theory in the context of 2+1-dimensional classical field theory \re{AG}. These
results were obtained for strong transitions, in which the nucleation barrier
$B$ was large. Nucleation was made possible due to the fairly large
temperatures
used in the simulations. (Recall that the decay time is proportional to
${\rm exp}(B/T)$.)

In order to study how the weakness of the transition will affect its dynamics,
the homogeneous part of the free-energy density (the effective potential
to some order in perturbation theory) is written as follows
\beq
U(\p ,T)={a\over 2}\left (T^2-T_2^2\right )\p^2-
{{\alpha}\over 3}T\p^3 +{{\l}\over 4}
\p^4 \:.
\label{e:freen}
\eeq
With this parameterization, the free-energy resembles the finite-temperature
effective potential
used in the description of the electroweak transition, where $\alpha$ is
determined by the masses of the gauge bosons and $T_2$, the spinodal
instability temperature, is related to the zero-temperature
mass of the Higgs boson \re{KRS}. Here, we will not be concerned with the
limits of validity of this effective potential in the context of electroweak
transitions. The goal is to explore the possible dynamics of a
transition with free energy given by Eq.~ \ref{e:freen}. This free-energy
is also remarkably similar to the de Gennes-Ginsburg-Landau free energy used in
the study of the nematic-isotropic transition of liquid crystals \re{LIQCRYS}.
This transition is known to be weakly first-order; there is a
discontinuity in the order parameter, even though there is no release of latent
heat \re{STINSON}. In fact, departures from mean field estimates for the
correlation length were detected as the degeneracy temperature is approached
from above, signalling the presence of ``pseudo-universal phenomena'',
characterized by long-wavelength fluctuations observed by
light-scattering experiments.

In 2+1 dimensions, it proves convenient to introduce dimensionless variables,
${\tilde x}=x\sqrt{a}T_2,~{\tilde t}=t\sqrt{a}T_2,~X=\p/\sqrt{T_2},~\theta=
T/T_2$, so that we can write the Hamiltonian (free energy) as
\beq
{{H[X]}\over {\theta}}={1\over {\theta}}\int d^2{\tilde x}\left [
{1\over 2}\mid \bigtriangledown X\mid^2 + {1\over 2}\left (\theta^2 -1
\right )X^2 -{{
{\tilde \alpha}}\over 3}\theta X^3+{{{\tilde \l}}\over 4}X^4\right ] \:,
\label{e:hamilton}
\eeq
where ${\tilde \a}=\a/(a\sqrt{T_2})$, and ${\tilde \l}=\l/(aT_2)$. (From
now on the tildes will be dropped.)
For temperatures above $\theta_1=(1-\a^2/4\l)^{-1/2}$ there is only one
minimum at $X=0$. At $\theta=\theta_1$ an inflection point appears at
$X_{\rm inf}=\a\theta_1/2\l$. Below $\theta_1$ the inflection point separates
into a maximum and a minimum given by $X_{\pm}={{\a\theta}\over {2\l}}\left [
1\pm \sqrt{1-4\l\left (1-1/\theta^2\right )/\a^2}\right ]$. At the critical
temperature $\theta_c=(1-2\a^2/9\l)^{-1/2}$ the two minima, at $X_0=0$ and
$X_+$ are degenerate. Below $\theta_c$ the minimum at $X_+$ becomes the global
minimum and the $X_0$-phase becomes metastable. Finally, at $\theta=1$ the
barrier between the two phases disappears. Note that $\a^2 < 4\l$ for a
solution as described above.

In order to study numerically
the approach to equilibrium at a given temperature
$\theta$, the coupling of the order
parameter $X$ with the thermal bath will be modelled by a Markovian Langevin
equation,
\beq\label{langevin}
{\partial^2X\over\partial t^2} = \bigtriangledown^2X -
{\tilde \eta} {\partial X\over
\partial t}
- {\partial U(X,\theta) \over \partial X} + {\tilde \xi}(x,t)~~,
\eeq
where ${\tilde \eta}=\eta/\sqrt{a}T_2$ is the dimensionless
viscosity coefficient,
and ${\tilde \xi}=\xi/aT_2^{5/2}$ is the dimensionless
stochastic noise with vanishing mean, related to
$\eta$ by the fluctuation-dissipation theorem,
\beq\label{flucdis}
\langle \xi(\vec x,t)\xi( \vec x',t')\rangle =
  2\eta T\delta(t-t')\delta^2(\vec x - \vec x')~~.
\eeq
The viscosity coefficient was set to unity in all simulations.
Two important comments are in order. In principle
it should be possible to obtain a Langevin-like equation for the slower
long-wavelength modes from a microscopic
approach by integrating out the short-wavelength modes which have faster
relaxation time-scales. This programme
is quite complicated in the context of relativistic field theories
since dissipation is a two-loop effect and
progress has been slow \re{MICROS}. Recent results indicate that one
should expect corrections to the above equation, as noise will in general
be colored and multiplicative as opposed to additive as above \re{MICROSII}.
It is possible that these corrections will change time-dependent quantities,
such as equilibration time-scales and nucleation rates, although they should
not affect final equilibrium properties of the system, such as the fractional
volume in each phase, or its critical properties,
which are most important here.
In lack of a better understanding of the microscopic
dynamics of such systems, the above equation will be adopted here as a starting
point. A second important point is to note that as this equation will be
solved on a lattice, the lattice spacing works as an effective
hard-momentum cut-off. Therefore, the lattice formulation is an effective
coarse-grained formulation of the continuum theory and one should be
careful when mapping from the lattice to the continuum theory. If one is
to probe physics at short-wavelengths, renormalization countertems should
be included in the lattice formulation so that the results are cut-off
independent and a proper continuum limit is obtained.
This point was emphasized and renormalization counterterms
obtained for a temperature-independent potential
in the nucleation study of Alford and Gleiser in Ref. \re{AG}. Here, due
to the temperature dependence of the potential, the renormalization
prescription of Alford and Gleiser does not work. Instead, the lattice spacing
will be fixed at $\ell =1$. It turns out that in all cases of interest here
the mean-field correlation length $\xi^{-2} = U^{\prime \prime}
(X_0,\theta)$ (not to be confused with the random
noise) will be sufficiently larger than unity to justify this choice.

The Langevin equation was integrated using the fifth-degree Nordsiek-Geer
algorithm which allows for fast integration with high numerical accuracy
\re{NORD}. The time step used was $\delta t=0.2$, and the results discussed
here were obtained with a square lattice with $L=64$. (Comparison with
$L=40$ and $L=128$ produced negligible differences for our present purposes.)
No dependence of the
results was found on the lattice length, time-step,
random noise generator, and the random noise seed.

The strategy adopted was to study the behavior of the system given by Eq.
\re{e:hamilton} at the critical temperature when the two minima are
degenerate. The reason for this choice of temperature is simple. If at
$\theta_c$ most of the system is found in the $X=0$ phase then as the
temperature drops below $\theta_c$ one expects homogeneous nucleation to
work; the system is well-localized in its metastable phase. This is what
happens when a system is rapidly cooled below its critical temperature
(rapid quench) so that it finds itself trapped in the metastable state. The
large amplitude fluctuations which will eventually appear are the nucleating
bubbles.
If at $\theta_c$ one finds a large probability for the system to be in the
$X_+$-phase, then considerable phase-mixing is occuring and homogeneous
nucleation should not be accurate in describing the transition. Large amplitude
fluctuations are present initially in the system, before it is quenched to
temperatures below $\theta_c$.
For definiteness call the two phases the 0-phase and the +-phase. The
phase distribution of the system can be measured if the idea of fractional
area (volume in 3 dimensions) is introduced. As the field evolves according
to Eq. \re{langevin}, one counts how much of the total area of the lattice
belongs to
the 0-phase
with $X\leq X_-$ ({\it i.e.} to the left of the maximum),
and how much belongs to the +-phase with $X>X_-$ ({\it i.e.} to the right of
the maximum). Dividing by the total area one obtains the fractional area in
each phase, so that $f_0(t) + f_+(t)=1$, independently of $L^2$.

The system is prepared initially in the 0-phase, $f_0(0)=1$ and $f_+(0)=0$.
Thus, the area-averaged value of the order parameter, $\langle X\rangle (t)=
A^{-1}\int X dA$ is initially zero. The coupling with the thermal bath will
induce fluctuations about $X=0$. By keeping $\l=0.1$ fixed, the dependence of
$f_0(t),~f_+(t)$, and  $\langle X\rangle (t)$ on the value of
$\a$ can be measured. Larger values of $\a$ imply
stronger transitions. This is clear from the expression for $\theta_c$ which
approaches unity as $\a\rightarrow 0$. That is, for small $\a$ the critical
temperature approaches the spinodal temperature. (In the electroweak case, the
same argument applies, as what is relevant is the ratio $\a^2/\l$; $\a$ is
fixed but $\l$ increases as the Higgs mass increases.) The results are
shown in Figure 1 for several values of $\a$ between $\a=0.3$ and $\a=0.4$.
Each one of these curves is the result of an
ensemble average over 200 runs. The two important features here are the final
equilibrium fraction in each phase and the equilibration time-scale.

The approach to equilibrium can be fitted at all times to a slow exponential,
\beq
f_0(t) = \left(1 - f_0^{{\rm EQ}}\right ){\rm exp}\left [-\left (t/\t_{{\rm
EQ}}
\right )^{\sigma}\right ] + f_0^{{\rm EQ}}\:,
\label{e:equil}
\eeq
where $f_0^{{\rm EQ}}$ is the final equilibrium fraction and $\t_{{\rm EQ}}$
is the equilibration time-scale. In Table 1 the
values of $\t_{{\rm EQ}}$ and $\sigma$ are listed for different values of
$\a$.\\
Note that the slot for $\a=0.36$ is empty. For this value of $\a$ the
approach to equilibrium cannot be fitted at all times to a slow exponential;
at large times it must be fitted
to a power law,
\beq
f_0(t)\mid_{\a=0.36} ~~\propto ~~ t^{-k} \:,
\label{e: powerlaw}
\eeq
where $k$ is the critical exponent controlling the approach to equilibrium.
An excellent fit is obtained for $k = 0.25 ~(\pm 0.05) $,
as shown in Figure 2.

\vspace{1.cm}

{\bf Figure 1:} The approach to equilibrium for several values of $\a$. \\

\vspace{1.cm}

{\bf Table 1}:  The values of the equilibration time-scales
and the exponents for
the exponential fit of Eq. \ref{e:equil}
for several values of $\a $. Also shown are the
equilibrium fractions $f_0(\theta_c)$ and $f_+(\theta_c)$. Uncertainties are
in the last digit.

\begin{center}
\begin{tabular}{|c|c|c|c|c|}\hline\hline
$\a $ & $\t_{{\rm EQ}}$ & $\sigma $ & $f_0(\theta_c)$ & $f_+(\theta_c)$ \\
\hline\hline
 0.30 & 21.0 & 0.80 & 0.505 & 0.495\\
 0.33 & 40.0 & 0.80 & 0.514 & 0.486\\
 0.35 & 75.0 & 0.60 & 0.525 & 0.475\\
 0.36 & --   & --   & 0.580 & 0.420\\
 0.37 & 25.0 & 0.65 & 0.800 & 0.200 \\
 0.38 & 15.0 & 0.80 & 0.870 & 0.130\\
 0.40 & 5.0  & 1.0  & 0.937 & 0.063\\
\hline
\end{tabular}
\end{center}

\vspace{1.0 cm}

The fact that there is a slowing down of the dynamics for $\a=0.36$
is indicative of the presence of a phase transition near $\a\simeq 0.36$.
This transition reveals itself in a striking way if we define as an order
parameter the equilibrium fractional difference $\Delta F_{{\rm EQ}}$,
\beq
\Delta F_{{\rm EQ}} = f_0^{{\rm EQ}} - f_+^{{\rm EQ}} \:.
\label{e:fbar}
\eeq


\vspace{1.cm}

{\bf Figure 2}: Fitting $f_0(\theta_c)$ to a power law at large times for
$\a=0.36$.\\

In Figure 3  $\Delta F_{{\rm EQ}}(\theta_c)$
is plotted as a function of $\a$. Clearly, there is a sharp transition in the
behavior of the system around $\a=\a_c\simeq 0.36$. For $\a < \a_c$ the
fractional area occupied by both phases is practically the same at 0.5. There
is considerable mixing of the two phases, with the system unable to
distinguish between them. One may call this phase the symmetric phase with
respect to the order parameter  $\Delta F_{{\rm EQ}}$. For $\a > \a_c$ there
is a clear distinction between the two phases, with the +-phase being sharply
suppressed. This may be called the broken-symmetric phase. It is clear from the
behavior of  $\Delta F_{{\rm EQ}}$ with $\alpha$ that this is a second order
phase transition. The rounding of the curve about $\alpha_c$ is due to
the finite size of the lattice. This curve is to be compared with the
behavior of the magnetization as a function of the temperature in Ising
models. $\alpha$ plays the r\^ole of ``inverse'' temperature; small $\alpha$
means high temperature, when symmetry is restored.

As a consequence of this behavior
a very clear distinction between a strong and weak transition is possible.
A strong transition has $\a > \a_c$ so that the system is dominated by the
0-phase at $\theta_c$. For a weak transition neither phase clearly dominates
the system, and as argued above, the dynamics should be quite different from
the usual nucleation mechanism. A mixing of the two phases occurs as the
system approaches the critical temperature from above.

It should be
possible to parameterize the behavior of  $\Delta F_{{\rm EQ}}$ in the
neighborhood of $\a_c$ with a power law,
\beq
\Delta F_{{\rm EQ}} \propto (\a - \a_c)^{\beta} \:,
\label{e:fbarcrit}
\eeq
where $\beta$ is the critical exponent controlling the behavior of
$\Delta F_{{\rm EQ}}$ near $\a_c$. The determination of this critical exponent
and of a more precise value of $\a_c$
will be left for a future publication as it involves a detailed analysis
of finite-size scaling \re{NUMISING}.


\vspace{1.cm}
{\bf Figure 3}:  The fractional equilibrium population difference $\Delta
F_{{\rm EQ}}$ as a function of $\a$. \\

In order to understand the reason for the sharp change of behavior
of the system near $\a_c$, in Figure 4 the equilibrium
area-averaged order paramemeter $\langle X\rangle_{{\rm EQ}}$ and the
inflection point $X_{{\rm inf}} = {{\a\theta}\over {3\l}}\left [1-\sqrt{1-
3\l(1-1/\theta^2)/\a^2}\right ]$, are shown as a function of $\a$. Also shown
is the rms amplitude of correlation-size fluctuations $X_{{\rm rms}}^2 =
\langle X^2\rangle_T - \langle X\rangle^2_T $, where
$\langle \cdots \rangle_T$
is the normalized thermal average with probability distribution $P[X_{sc}
] ={\rm
exp}\left [-F[X_{sc}]/\theta\right ]$. $F[X_{sc}]$ is the free energy of a
gaussian-shape subcritical fluctuation.
For details see Ref. \re{GRII}. It is clear
from this Figure that the transition from weak to strong occurs as
$\langle X\rangle_{{\rm EQ}}$ drops below  $X_{{\rm inf}}$.
This can be interpreted as an effective Ginzburg criterion
for the weak-to-strong transition. It matches quite
well the fact that the critical slowing down occurs for $\a\simeq 0.36$.
This result is in qualitative agreement with the study of Langer {\it et al.}
contrasting the onset of nucleation vs. spinodal decomposition
for binary fluid and solid solutions
\re{LANGER II}, where it was found that the transition between the two
regimes occurs roughly at the spinodal ({\it i.e.} at the inflection point).
Note, however, that here we are dealing with relativistic field theories, while
the work of Langer and collaborators had to do with phenomenological models
of phase transitions for systems with conserved order parameter. These systems
will typically have slower dynamics than the field theories of interest in
cosmology. Also, for Langer and his collaborators, as in most applications in
condensed matter physics, the dynamics is studied {\it after} the system is
quenched (rapidly cooled) to below its critical temperature. In cosmology, the
cooling is provided by the expansion rate of the universe.

Even though $X_{{\rm rms}}$ drops below  $X_{{\rm inf}}$ for a smaller
value of $\a$, being a much less computer intensive quantitity to obtain,
it should serve as a rough indicator of the weak-to-strong transition.

\vspace{1.cm}
{\bf Figure 4}:  Comparison between area-averaged field and location of the
inflection point as a function of $\a$. Also shown are the location of the
barrier, $X_{{\rm MAX}}$, and the rms fluctuation $X_{{\rm rms}}$.  \\

To summarize the results of this Section, the distinction between a weak and
a strong first order transition can be made quantitative by studying the
behavior of the effective free energy {\it at the critical temperature}.
A clear
change in behavior occurs at a critical value of the parameter used to
define the ``strength'' of the barrier separating the two phases. Here,
the cubic coupling was chosen as an example. For values of the parameter
larger than the critical value, the system is mostly in the symmetric
phase and the transition should be well described by homogeneous nucleation
theory. For values of the parameter below its critical value, the system
is in a mixed phase and no phase is preferred. (Of course this assumes that
the system is being cooled down slowly compared to its typical fluctuation
rates. If we rapidly
quench the system to below its critical temperature and then keep cooling it
further,
it will not reach the mixed-phase state. In cosmology this would correspond to
a {\it very} early transition, when the universe's expansion
rate is relatively fast compared
to typical fluctuation rates in the system.)
In this case, the transition
will proceeed by a mechanism closer to spinodal decomposition, although
more studies are needed to obtain a clear picture of the approach to
equilibrium
in this case. The transition between weak  and strong is itself a
(second order) phase
transition, with the equilibrium fractional population difference playing the
r\^ole of the order parameter.

Finally let me stress that these results are not particular to 2+1 dimensions.
Indeed, recent work on 3+1 dimensions produced qualitatively identical
results. There is a second order transition  for a critical value of
$\alpha$ which for a given value of $\lambda$,
will be lower than the 2+1 dimensional value. (Systems fluctuate more in two
dimensions.) In
order to make (some) contact with the electroweak transition, we should
investigate how $\lambda$ (related to the Higgs mass)
varies for fixed $\alpha$ (related to the mass of the gauge bosons). Even
though we investigated a model with a real scalar field, this should give us
an indication of the transition from weak to strong in the electroweak case
as well \re{WEAKII}.

\vspace{0.5cm}

\noindent{\bf IV. Subcritical Bubbles:
A Simple Approach to Nonequilibrium Dynamics}\\

Now I finally, and briefly, turn to the subject of subcritical bubbles, as
a possible method to study the approach to equilibrium in cosmology (and
in the laboratory). Consider a system described by an effective
(coarse-grained)
free-energy density as discussed above, for example. If we prepare the system
in the symmetric phase (or, in the example above --with no symmetry-- at
$\langle \phi \rangle = 0$),
at any temperature there will be fluctuations which will probe the other phase.
These fluctuations will be suppressed by a Boltzmann factor. The larger the
amplitude of the field and the larger the volume of the
fluctuation, the larger the suppression. Also, once they appear, they will
disappear, unless the system is below the critical temperature and
they happen to be larger than the critical fluctuations for nucleation.
However, at a given temperature, {\it they will always be there.}
The whole discussion and results of the previous Section offer
convincing evidence that systems fluctuate about their
equilibrium values, sometimes quite dramatically so. (The nice thing about
numerical --and most-- experiments is that they are reproducible.
You can, if so disposed, always convince
yourself that there is a clear distinction between ``weak'' and strong
first order transitions.)
The subcritical bubbles method was proposed in order
to obtain a semi-analytic description of these fluctuations so that we
can examine their importance. In the light of the previous results, for
strong transitions they should be irrelevant, while for weak transitions they
are crucial. In passing, I note that subcritical fluctuations of the
broken-symmetric phase within the symmetric phase have been observed
{\it above the critical temperature} in the isotropic-nematic liquid crystal
transition \re{LIQCRYSRELAX}. Nematic fluctuations within
the isotropic phase were identified
and their relaxation time measured, in order to study departures from
mean-field theory predictions. As expected, only in the neighborhood of the
critical temperature substantial departures from mean-field were observed.

Expanding on the GK approach, Gelmini and Gleiser (GG) obtained a kinetic
equation that incorporates both
the shrinking of the subcritical bubbles, and their
possible ``destruction'' by thermal noise \re{GG}. If $n(R,t)$ is the number
density of subcritical bubbles of radius $R$ of the broken-symmetric phase
within the symmetric phase, the rate equation is
\begin{eqnarray}
\label{eq:KIN}
{{\partial \n}\over {\partial t}}=-{{\partial \n}\over {\partial R}}
\left ({{dR}\over {dt}}\right )+\left ({{V_0}\over V}\right )\Gamma_{0
\rightarrow +}(R)  \nonumber\\
 - \left ({{V_+}\over V}\right )\Gamma_{+\rightarrow 0}(R)
- \left(\frac{V_+}{V}\right) \Gamma_{TN} (R)
\end{eqnarray}
Here, $\Gamma_{0\rightarrow +}(R)$ ($\Gamma_{+\rightarrow 0}(R)$)
is the rate per unit volume for
the thermal nucleation of a bubble of radius $R$ of phase $\f=\f_+$ within
the phase $\f=0$ (phase $\f=0$ within
the phase $\f_+$). $\Gamma_{TN}  (R) \simeq a
T/\frac{4}{3} \pi R^3$ is the (somewhat ad hoc)
expression used for the thermal destruction
rate, with $a$ a constant. Also, $V_+$
must be understood as the volume of the (+)--phase in bubbles of radius $R$
{\it only}, since we are following the evolution of $\n$.

In order to obtain analytical solutions of this equation, GG solved it
only for temperatures just below the temperature when the broken-symmetric
minimum appears which ($\theta_1$ in the model above),
of course, is above the critical temperature.
For this temperature, most of the system will always be in the symmetric phase
and we can write $V_0/V\simeq 1$. Another important assumption is that
most of the subcritical bubbles will be of correlation volume. This is due
to the fact that larger fluctuations will be suppressed, while smaller
fluctuations are inconsistent with the coarse-graining procedure.
With these assumptions, it is possible to solve the kinetic equation and
obtain two crucial quantities; the equilibration time-scale typical for
each of the processes that suppress subcritical bubbles (shrinking, thermal
nucleation of a subcritical bubble of the symmetric phase inside a region
of broken symmetric phase, and thermal destruction), and the equilibrium
number density of these bubbles. This way we can distinguish which process
is the dominant process for the suppression of subcritical bubbles for
different parameters of the free energy density. Applying this formalism
to the 1-loop electroweak potential, GG showed that for Higgs masses above
55 GeV or so, considerable phase mixing is occurring even for temperatures
above the critical temperature. Thus, the subcritical bubble picture is
in excellent qualitative agreement with the numerical results described in the
previous Section; for weak enough transitions we should expect substantial
departures from the usual vacuum decay mechanism.
\vspace{0.5cm}

\noindent{\bf V. Concluding Remarks}\\

In this talk I discussed some of the issues related to the dynamics of weak
vs. strong first order phase transitions. As I hope was made clear,
if indeed the electroweak phase transition is ``weak'' in the sense defined
here, novel aspects of nonequilibrum dynamics will have to be taken into
account when dealing with the computation of the net baryon number generated
in a given model. Taken at face value, the results here may be bad news
for electroweak baryogenesis in the context of the standard model. Even if
lattice computations show that for Higgs masses of 80 GeV or so the transition
is still strong, the Higgs may weigh much more than that. This being the
case, there will always be a regime in which phase mixing will occur and
nucleation theory will fail. On the other hand, apart from hand-waiving
arguments, it is not clear that the domain coarsening dynamics that will take
place in a weak transition will not produce a net baryon number. At present,
we simply do not know enough about the nature of the approach to equilibrium
to decide on this issue, or set it aside. If baryon number is to be generated
in
extensions of the standard model, then the results here will provide a useful
constraint in the usually large parameter space of these models. In order
to have a strong enough transition, the Ginzburg-like criterion discussed
here must be satisfied, and at least one parameter of the model may be
eliminated this way. If the transition in these models is weak enough,
one should again expect departures from the standard vacuum decay estimates for
bubble nucleation rates. This is due to the inescapable conclusion that
hot systems fluctuate, and if they can these fluctuations will produce some
dramatic effects. Critical opalescence in nematic liquid crystals is but one
example of these ``pre-transitional'' phenomena in nature. It may well be
that they will also be of crucial importance in cosmological phase transitions.

The possibility of generating the baryon number of the universe during the
electroweak phase transition is a tremendous challenge to present research
in the cosmology/high energy physics interface. We most probably must
invoke physics beyond the standard model to get enough CP violation, we must
tackle hard problems related to infrared divergences in gauge theories, and,
last but not least, we must understand the nonequilibrium aspects of phase
transitions in the context of field theories in a cosmological background.
Judging from what happened during the past 5 years or so, and by the number
of people working in this topic, progress will keep coming fast. Maybe in
another 5 years, we should all get together again (hopefully in
Sintra) to see how far we managed to get.

\vfill\eject

\noindent{\bf Acknowledgements}
\vspace{0.5cm}

I am grateful to my collaborators Rocky Kolb and Graciela Gelmini, as well
as M. Alford and R. Ramos for the many long discussions on bubbles and
phase transitions. This work is partially supported by a
National Science Foundation grant
No. PHYS-9204726.

\vspace{1.cm}

\noindent{\bf References}
\begin{enumerate}

\item\label{BD}  E. Becker and W. D\"oring, Ann. Phys. (Leipzig) {\bf 24},
719 (1935).

\item\label{LANGER}  J. W. Cahn and J. E. Hilliard,
             J. Chem. Phys. {\bf 31}, 688 (1959);  J. S. Langer,
Ann. Phys. (NY) {\bf 41}, 108 (1967);
{\it ibid.} {\bf 54}, 258 (1969)

\item\label{CH}
J. D. Gunton, M. San Miguel,
and P. S. Sahni, in
{\it Phase Transitions and Critical Phenomena}, edited by C. Domb and J. L.
Lebowitz (Academic, London, 1983), Vol. 8.

\item\label{VKO}  M. B. Voloshin, I. Yu. Kobzarev, and L. B. Okun',
        Yad. Fiz. {\bf 20}, 1229 (1974)
        [Sov. J. Nucl. Phys. {\bf 20}, 644 (1975) ].

\item\label{COLEMAN} S. Coleman, Phys. Rev. {\bf D15}, 2929 (1977); C. Callan
and S. Coleman, Phys. Rev. {\bf D16}, 1762 (1977).

\item\label{LINDE}   A. D. Linde, Nucl. Phys. {\bf B216}, 421 (1983);
[Erratum:
{\bf B223}, 544 (1983)].

\item\label{FINITETDECAY}
L. P. Csernai and J. I. Kapusta, Phys. Rev. {\bf D46},
1379 (1992); M. Gleiser, G. Marques, and R. Ramos, Phys. Rev. {\bf D48},
1571 (1993); D. Brahm and C. Lee, Phys. Rev. {\bf D49}, 4094 (1994).

\item\label{CPT} E. W. Kolb and M. S. Turner, {\it The Early Universe},
Addison-Wesley (1990).

\item\label{KRS} V. A. Kuzmin, V. A. Rubakov,
and M. E. Shaposhnikov, Phys. Lett. {\bf B155}, 36 (1985).
For a recent review see, A. Cohen, D. Kaplan, and A. Nelson, Ann. Rev.
Nucl. Part. Sci. {\bf 43}, 27 (1993).

\item\label{BRANDEN} R.H. Brandenberger and A.-C. Davis, Phys. Lett.
{\bf B308}, 79 (1993).

\item\label{THICK} B. Liu, L. McLerran, and N. Turok, Phys. Rev.
{\bf D46}, 2668 (1992); M. Dine and S. Thomas, Santa Cruz preprint No.
SCIPP 94/01.

\item\label{EWPERTURB} P. Arnold and O. Espinosa, Phys. Rev. {\bf D47}, 3546
(1993); M. Dine, P. Huet, R.G. Leigh, A. Linde, and D. Linde, Phys. Rev.
{\bf D46}, 550 (1992); C.G. Boyd, D.E. Brahm, and S. Hsu, Phys. Rev.
{\bf D48}, 4963 (1993); M. Quiros, J.R. Spinosa, and F. Zwirner, Phys. Lett.
{\bf B314}, 206 (1993); W. Buchm\"uller, T. Helbig, and D. Walliser,
Nucl. Phys. {\bf B407}, 387 (1993); M. Carrington, Phys. Rev. {\bf D45},
2933 (1992).

\item\label{EPSILON} P. Arnold and L. Yaffe, Phys. Rev. {\bf D49},
3003 (1994);
M. Gleiser and E.W. Kolb,  Phys. Rev. {\bf D48}, 1560 (1993).

\item\label{EWLATTICE} K. Farakos, K. Kajantie, K. Rummukainen, and M.
Shaposhnikov, CERN preprint No. TH.7244/94, May 1994;
K. Kajantie, K. Rummukainen, and M. Shaposhnikov,
Nucl. Phys. {\bf B407}, 356 (1993); B. Bunk, E.-M. Ilgenfritz, J. Kripfganz,
and A. Schiller, Phys. Lett. {\bf B284}, 372 (1992); Nucl. Phys. {\bf B403},
453 (1993)

\item\label{GR} See, e.g. the work of Gleiser, Marques, and Ramos in
Ref. \ref{FINITETDECAY}; P. Arnold and L. McLerran, Phys. Rev. {\bf D36},
581 (1987); S. Dodelson and B. Gradwohl, Nucl. Phys. {\bf B400}, 435 (1993).

\item\label{GK} M. Gleiser and E. W. Kolb, Phys. Rev. Lett. {\bf 69},
1304 (1992); Phys. Rev. {\bf D48}, 1560 (1993);  M. Gleiser, E. W. Kolb, and
R. Watkins, Nucl. Phys. {\bf B364}, 411 (1991).

\item\label{CRITICS}  M. Dine, P. Huet, R.G. Leigh, A. Linde,
and D. Linde, Phys. Rev.
{\bf D46}, 550 (1992); G. Anderson, Phys. Lett. {\bf B295}, 32 (1992).

\item\label{GG} G. Gelmini and M. Gleiser, in press, Nucl. Phys. B.

\item\label{WEAK} M. Gleiser, Dartmouth preprint No. DART-HEP-94/01.

\item\label{NUMISING} D. P. Landau in {\it Finite Size Scaling and
Numerical Simulations of Statistical Systems}, edited by V. Privman
(World Scientific, Singapore, 1990).

\item\label{AG} M. Alford and M. Gleiser, Phys. Rev. {\bf D48}, 2838 (1993);
O. T. Valls and G. F. Mazenko, Phys. Rev. {\bf B42},
6614 (1990).

\item\label{LIQCRYS}
S. Chandrasekhar,
{\it Liquid Crystals}, (Cambridge University Press, Cambridge
[Second edition, 1992]).

\item\label{STINSON} T. W. Stinson and
J. D. Litster, {\it Phys. Rev. Lett.} {\bf 25}, 503 (1970); H. Zink and
W. H. de Jeu, {\it Mol. Cryst. Liq. Cryst.} {\bf 124}, 287 (1985).

\item\label{MICROS}  K. Chou, Z. Su, B. Hao and L. Yu, Phys. Rep.
{\bf 118}, 1 (1985);
R. R. Parwani, Phys. Rev. {\bf D45}, 4695 (1992);
S. Jeon, Phys. Rev. {\bf D47}, 4586 (1993).

\item\label{MICROSII} M. Gleiser and R. Ramos, Dartmouth preprint No.
DART-HEP-93/06;  B. L. Hu, J. P. Paz and Y. Zhang, in {\it The Origin of
Structure in the Universe}, Ed. E. Gunzig and P. Nardone
(Kluwer Acad. Publ. 1993);  D. Lee and D. Boyanovsky, Nucl. Phys. {\bf B406},
631 (1993);  S. Habib, in {\it Stochastic Processes in Astrophysics},
Proc. Eighth Annual Workshop in Nonlinear Astronomy (1993).

\item\label{NORD} A. Nordsiek, Math. Comp. {\bf 16}, 22 (1962).

\item\label{GRII} M. Gleiser and R. Ramos, Phys. Lett. {\bf B300}, 271 (1993).

\item\label{LANGER II} J. Langer, Physica {\bf 73}, 61 (1974);
J. Langer, M. Bar-on, and H. Miller, Phys. Rev. {\bf A11}, 1417 (1975).

\item\label{WEAKII} J. Borrill and M. Gleiser, in progress.

\item\label{LIQCRYSRELAX} F. W. Deeg and M. D. Fayer, Chem. Phys. Lett.
{\bf 167}, 527 (1990);  J. D. Litster and T. W. Stinson, J. App. Phys.,
{\bf 41}, 996 (1970).

\end{enumerate}

\end{document}